# ReducedCBT and SuperCBT
## Two New and Improved Complete Binary Tree Structures

**Mevlut Bulut**
Birmingham, Alabama, USA
mevlutbulut@yahoo.com

**Abstract:** Between the leaves and the nodes of a complete binary tree, a separate parent-child-sister hierarchy is employed independent of the parent-child-sister hierarchy used for the rest of the tree. Two different versions of such a local hierarchy are introduced. The result of the first proposed hierarchy is a faster and smaller footprint, while the second one provides the size variation functionality without a significant computational overhead. This novel approach brings considerable memory gains and performance boosts to the complete binary tree based algorithms.

## 1 Introduction

Modern computers have different levels of memory fields, e.g. cache memory, main memory, and disk drive unit. These different memory fields have different access parameters. The performance of an algorithm is heavily dependent on memory usage characteristics of that algorithm. When the data substrate fits inside the fastest memory field, namely level-1 cache memory, the performance depends on the conciseness of the code and the contained operations. As the tackled data floods into slower memory fields, the memory operations become a hindrance to the efficiency of the algorithm. In order to design highly efficient general purpose algorithms, some researchers used the cache-oblivious paradigm [2] to reduce the quantity of memory operations through the slower memory fields, while some others [3] suggested architecture sensitive data layouts in order to ease the bottlenecks for slower memory field operations. In this article however, a different technique is proposed; substitution of slower (far) memory field operations by algebraic operations, which do not require reaching far memory fields at all.

Complete Binary Tree (CBT) structure has been a useful tool for applications like priority queues and sorting algorithms since the date it was introduced [1, 4, 5]. CBT based algorithms demonstrate predictable performances, in other words; for a CBT based implementation, there is not much difference between the best case and the worst-case scenarios. Despite its prevalent use, growing and shrinking ability of CBT is not ideal. For example, when CBT is used in a sorting algorithm to sort an array of labeled keys , the size of the CBT can be fixed, in which case a dummy key(computer representation of either positive or negative infinity depending on the selection rule used) is used to replace the removed keys [5]. This approach has two disadvantages. First, no matter which dummy key value is used, that value puts a restriction to the array of elements being sorted not to have it. Second, although the number of unsorted elements goes down as sorting takes place, the required number of comparisons to find the successor element does not go down at all. This is because the emptied leaves are loaded with dummy keys to be used in subsequent comparisons during the update operations. To overcome these disadvantages, we can let the CBT physically shrink as the elements are dismissed from the tree. In this case, we can use auxiliary arrays to remap the indices to leaf nodes in order to tackle the unsystematic displacement of keys [6] as the tree shrinks. This method eliminates the above-mentioned disadvantages, but introduces two new ones: The first is the necessity of extra memory resources for the helper arrays (a total of $3N$ integer space, $N$ is the number of keys). The second is the increased cost of memory operations for this index redirection through the helper arrays.

A fast priority queue (PQ) implementation with size varying functionality is very desirable for event-driven particle simulations (EDPS) in cases when the number of simulated particles dynamically changes because of nucleation, coagulation, coalescence, or particle flow in and out of the system [7]. In an EDPS, time evolution of a particle population is rendered through consecutive discreet events such as binary collisions or cell crossing events. The CBT structure is already proven as the best structure to implement a PQ algorithm to be used with an EDPS [6, 8], mainly thanks to its simplicity and its working performance almost independent of the probability distribution of the incoming priority key values[1] (this point will further be addressed in the *Results and Discussion* section of this article). Nevertheless, when the size varying functionality is added to a CBT based PQ in the fashion Mauricio Marin suggested [1], the implementation requires a 3N auxiliary integer space together with the overhead of the index redirection through the helper arrays.

The main motivation of this research was to add size varying functionality to CBT based PQs [1, 4, 5] without incurring neither extra memory usage nor considerable performance degradation. In the modified CBT structures presented by this article, separate local parent-child-sister hierarchies are utilized instead of using extra helper arrays. ReducedCBT section explains how to construct a constant size CBT using only $N$ integer space, and the section after that focuses on how to improve this idea in order to acquire size variation functionality.

In a PQ, we use the key values corresponding to distinguished events in order to find the highest priority event. Therefore, keys are accompanied by event IDs. Key values and event IDs can be given as tuples or as separate arrays. In the case they are given in separate arrays, every key and its corresponding event ID should have the same array index. When the term 'key index' is used, we are talking about the place of the key in the array of keys used by the PQ. However, when the term 'event ID' is used, we are talking about a label of a specific event which we are manipulating its representative key. There may or may not be a relationship between the key indices and the corresponding event IDs. In terms of this relationship, three different cases can be identified:

1. Event IDs and key Indices are identical: Key indices are enough to identify the corresponding events. PQ applications for event-driven simulations with fixed number of items are good examples for this case.

2. There is a simple relationship between Event IDs and key Indices, such as a constant difference: Key indices can still be used to identify the corresponding events trough a simple algebraic manipulation.

3. Event IDs and key Indices are not related: A separate integer array is used to store the corresponding event IDs of the keys. A classic example for this case is a PQ application used for initial runs during an external sorting application. The winner keys are replaced by new keys with different IDs. Another good example for this case is a local PQ for the current day events of a calendar queue. In this case, event IDs will be completely random.

In the next section, we will focus on fixed size CBT structures and for simplicity, we will assume that the first case is true, but the ideas can easily be applied to other two cases.

## 2    ReducedCBT

During the construction of a fixed size CBT, we load the key indices into the leaf nodes. [Fig. 1-(a)] shows a CBT with 11 keys. To find the leaf node index $n$ for a given key index $i$, we simply add the number of total keys to $i$: $n \leftarrow i + N$; The way of finding parent and sister nodes suggested by [1] is as follows:

        $p \leftarrow n/2$;    // $p$ represents parent node.
        $l \leftarrow 2*p$;    // $l$ represents left sister node.
        $r \leftarrow 2*p +1$; // $r$ represents right sister node.

This is a top-down approach, the parent node is found first and then, left and right child nodes are calculated. [5] also suggests that top-down approach is often considerably simpler. Yet, in this article, a simpler bottom-up approach is suggested instead. This point is an important factor for the overall performance of the tree, also is a precursor for the new ideas introduced by this article. Using $s$ for sister node and $p$ for parent node, the suggested bottom-up approach can be expressed as follows:

        $s \leftarrow n$ XOR $1$; n is exclusive OR'ed with one.
        $p \leftarrow n >> 1$    ; n is right shifted once.

Here is why this approach is superior to the top-down approach: During an update operation, we compare the keys (hosted by sister nodes), and register the winner into the parent node. At that point, we know the content of that parent node. For the next iteration, all we need is to reach the sister node and the parent node of that node. This means, at every level of the tree, we need to reach only two new nodes; one of them is the new sister node for reading, the other one is the new parent node for writing. In the case of top-down approach however, new parent node is reached for writing, but both of its children nodes are reached for reading their contents. One of the descendent nodes is the previous parent node and its content is already known, therefore re-accessing that node for reading purpose is futile, and is avoided in the bottom-up approach suggested here. This saves a considerable amount of time during the update operations.

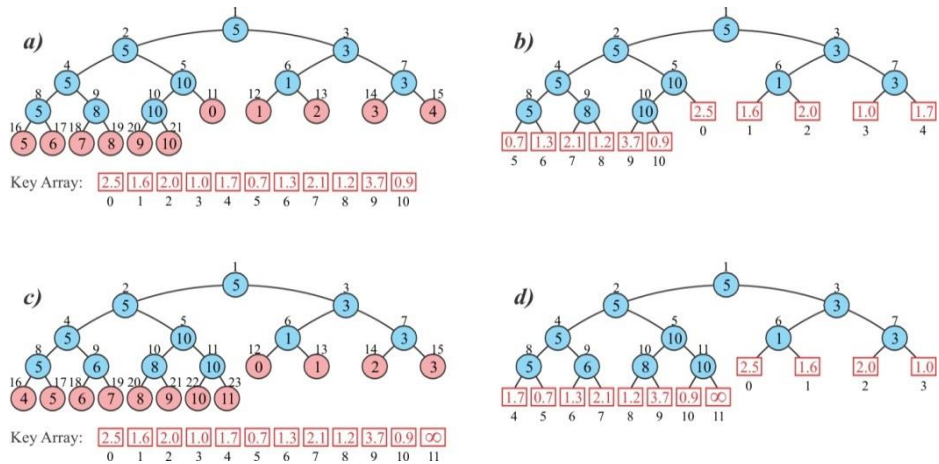

*Figure 1: (a) Initial placement of leaves for a CBT structure with 11 leaves. 2N integer space is used to form internal and leaf nodes. (b) The same tree using the same internal nodes but instead of using leaf nodes, the given key array is directly used as the leaves. (c) and (d) are similar to (a) and (b) except an even number of keys is used.*

In order to calculate the leaf node for a given key index, we just add the number of total keys to the key index as explained above. This simple relation suggests that we might be able to use a different parent-child-sister hierarchy at the bottom level to get rid of the leaf nodes. [Fig. 1-(b)] shows the implementation of this idea. The key array is directly used as the leaves. To test the idea, let us find the sister key "$s$" and the parent node "$p$" for a given key index "$i$".

$$s \leftarrow ((i+N) \text{ XOR } 1) - N;$$
$$p \leftarrow (i+N) / 2;$$

Here, by adding "$N$" to "$i$", we get a virtual leaf node index. In order to find "$s$", we subtract "$N$" from the result of the exclusive OR operation to convert back from virtual node index to key index. As an example, let us find the sister key and the parent node for the first key ($i=1$).

$$s \leftarrow ((1+11) \text{ XOR } 1) - 11 = 2;$$
$$p \leftarrow (1+11) / 2 = 6;$$

These values for $s$ and $p$ can be verified by examining [Fig. 1-(b)]. What happens when $i=0$?

$$s \leftarrow ((0+11) \text{ XOR } 1) - 11 = -1;$$
$$p \leftarrow (0+11) / 2 = 5;$$

Here $p$ value comes out as correct, but the sister key index turns out to be negative. It does not work. This is because the sister of the zeroth key is not another key but a tree node; there is a local irregularity there. This problem can be solved by using conditionals, but this will increase the computational overhead. As it can be seen from [Fig. 1-(c) and (d)], this problem does not occur when the number of keys is an even number.

$$i = 0, N = 12;$$
$$s \leftarrow ((0+12) \text{ XOR } 1) - 12 = 1;$$
$$p \leftarrow (0+12) / 2 = 6;$$

$$i = 1, N = 12;$$
$$s \leftarrow ((1+12) \text{ XOR } 1) - 12 = 0;$$
$$p \leftarrow (1+12) / 2 = 6;$$

It can be checked from [Fig. 1-(d)] that zeroth and first keys are sisters to each other and the parent node index is six. Because we are dealing with fixed size CBTs in this section, we can easily make the number of total keys even by adding a joker key in cases it is given as an odd number. Additionally, when $N$ is an even number, the result of $(i+N)$ XOR 1 is not affected by $N$, because the last bit of $(i+N)$ is defined by the last bit of $i$ when $N$ is even. Therefore, we do not have to add $N$ before the bitwise XOR and subtract it afterwards. Then the proposed local hierarchy can be expressed as:

$$s \leftarrow i \text{ XOR } 1;$$
$$p \leftarrow (i+N) / 2;$$

The following is the summary of the simplified local parent-child-sister hierarchy for the leaves and the pseudo code for a full update operation. A full update operation is defined as starting from a leaf node, reevaluating the winner element between the sister-nodes, and going up to the root, while a partial update proceeds up to a certain level where further comparisons will not change any of the above node contents:

```
//Update the value of the i'th key. //Selection rule: Smaller wins.
Integer N; //must be an even number.
Integer s; // index of the sister key.
Integer p; //index of the parent node.
//CBT[]: an integer array of size N, to be used as internal nodes.
//key[]: an array of size N to store the keys.

    s← i XOR 1;      //proposed local hierarchy: sister key index← given key index XOR 1;
    p← (i+N) /2;     //proposed local hierarchy : parent node ← (given key index + N) /2;
    Loop (until p becomes equal to zero)
        if (key[s] < key[i])  i← s;//variable i will hold the index of the winner key.
        CBT[p] ← i; //register the winner in the parent node.
        s← CBT[p XOR 1]; //global hierarchy: sister node ← current node XOR 1;
        p← p /2;            //global hierarchy: parent node ← current node /2;
    EndLoop
```

In order to assess the efficiency of the explained structure, the comparison section will contain a PQ implementation based on this structure under the name 'ReducedCBT'.

## 3    SuperCBT

If we apply a shrink-operation to the above-described ReducedCBT by removing the winner key [Fig. 2- (b)], the consecutive nature of the key indices will be violated. This will render the proposed local parent-child-sister hierarchy broken. Therefore, the local parent-child-sister hierarchy described in ReducedCBT section is not self-preserving under key removal operations. Now let us ask this question: Is there any self-preserving local parent-child-sister hierarchy that will enable usage of the keys as the leaves like in ReducedCBT, even under size-varying tree operations ( e.g. grow and shrink)? Yes, there are such hierarchies and one such hierarchy will be explained in this section: The CBT with this specific local hierarchy at the level of leaves will be called SuperCBT.

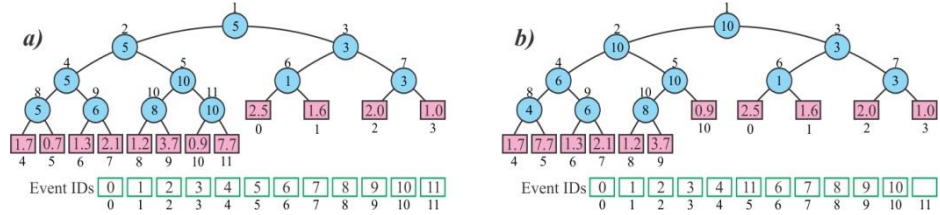

*Figure 2: a) shows an initialized ReducedCBT with 12 keys. b) depicts the resulting structure after deleting the winner key, moving the last key to the position of the deleted key and raising the sister of the last key to the position of its previous parent node. After this operation, the tree becomes dysfunctional since the consecutive nature of the leaves is violated. Unless the whole tree is reinitialized, it cannot be used as a ReducedCBT structure.*

[Fig. 3] depicts the proposed local parent-child-sister hierarchy for $N= 12$. Here is how we find the sister and fist parent node for a given key index $i$: First, we check if the given leaf is a right leaf or a left leaf. Keys with an index greater than or equal to $N/2$ are right keys and the rest of the keys are left keys. To find the sister (pairing keys are sisters of each other) of a right key, we find the number representing the position of the least significant set bit of the key index, and shift the key index to the right that many times. In other words, we are getting rid of

the rightmost block of consecutive zeros and the adjacent set bit to their left. For instance, the 10[th] key in [Fig. 3] is the right leaf and the least significant set bit of 10 (in binary *1010*) is the second digit. Therefore, we shift the binary representation of 10 to the right twice and obtain "*0010*" in binary, which is equal to two in decimal, and it is the key index of the left sister of 10[th] key. If the key at hand is a left key (index of it is smaller than *N*/2), we take the index of that key, first multiply it by two and add one. Then multiply it by two successively until the largest number smaller than the array size *N* is reached; that number will be the index of the right side sister key. Doing this in binary representation is very simple: index number of a key is shifted left once filling the new least significant bit by "1" and then it is shifted left as many as $n_{LS}$ (number of left shifts), filling the least significant bits by "0". For an index $i$, the number of left shifts, $n_{LS}$ is defined by the position of the most significant set bit of *maxIndex*/(2$i$ + 1) ratio and it is algebraically expressed as:

$$n_{LS} = int\left(Log_2 \frac{maxIndex}{2i+1}\right)$$

Here, *maxIndex* represents the number of active leaves minus one (*maxIndex* = *N* - 1), and *i* is the array index of the key for which we are finding the right sister key. The physical meaning of $n_{LS}$ is how many times we can double the value (2$i$ + 1) before the result exceeds *maxIndex*, namely how many down-left moves we should make in order to reach the leaf level of the CBT. The index of the first parent and the index of the right leaf are the same.

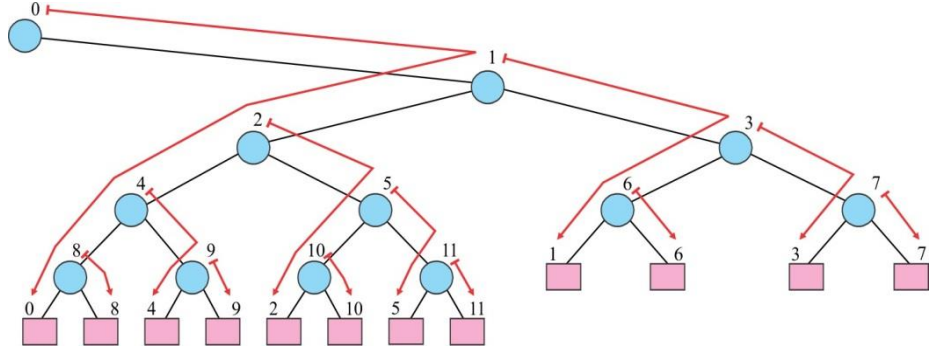

*Figure 3: The proposed local parent-child-sister hierarchy for the leaves of a CBT with 12 nodes. Zeroth node is shown only for the graphical representation of the proposed hierarchy.*

The following is the pseudo-code for the proposed parent-child-sister hierarchy between the lowest tree nodes and the leaves for the SuperCBT structure:

procedure FindParentAndSister (int winner, int parent, int sister)
    if ( winner  > (maxIndex/2) )  then   //it is a right leaf.
        parent←winner;
        int *LSSBp* ← FindLSSB_Position_of ( parent );
        // *LSSBp* stands for "least significant set bit position".
        sister← (right shift *parent* by *LSSBp*);
    else   //it is a left leaf.
        parent← 2*winner + 1;
        int MSSBp← FindMSSB_Position_of (*maxIndex*/*parent*);
        // MSSBp stands for "most significant set bit position".
        *parent*  ← (Left shift *parent*  by (MSSBp - 1));//if MSSBp starts counting from one.
        //*parent*  ← (Left shift *parent*  by MSSBp); //if MSSBp starts counting from zero.

```
        sister ← parent;
        if (sister >= maxIndex)  sister ← winner;
    endif
```

As an example of finding parent and sister indices for a key index, lets look at a tree of size 12, like the one in [Fig. 3]. If *key Index* is equal to one, which falls into the *else* case of the conditional block of FindParentAndSister method (hence it is a left key), the temporary value of *parent* will be three and the integer division, *maxIndex/parent,* will give three. The position of the most significant set bit of three is two (three in binary is *0011,* and counting from the right, the second bit of this binary number is its most significant set bit). Then the value *parent* will be shifted left once, rendering the value of *parent* to become six. [Fig. 3] shows that the parent node of leaf "1" is indeed "6". The little sorting example in [Fig. 4] graphically shows how the new structure works.

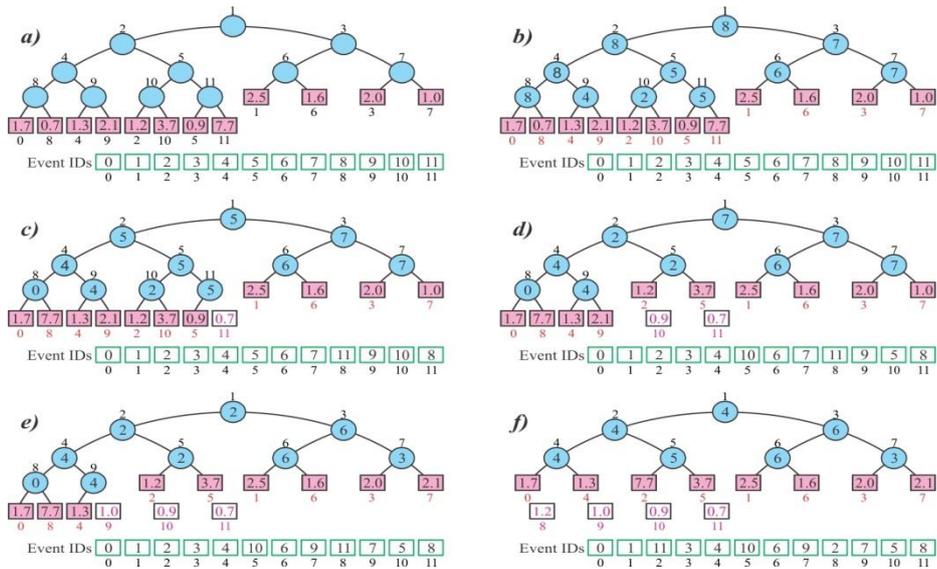

*Figure 4: A sorting example using the proposed SuperCBT structure. a) The initial placement of the keys for a SuperCBT with 12 leaves. From b) to f) show the successive applications of RemoveKey and Update operations depicting a sorting example using the proposed structure. Rectangular boxes are used to represent the given key array. The small numbers under the keys are the key indices. The keys excluded from the structure were left at their previous geographic sites but their links with the tree have been removed to show that they are not part of the SuperCBT anymore. In the end, the array called 'Event IDs" will have the IDs of the sorted keys. The ID of the smallest key will be at the end and the ID of the largest key will be at the beginning of the array.*

This structure can also be used in priority queue applications where no key removal is required. In applications where a key removal is required, the last tree node content is tested if it contains the index of the last key or not. If the index of the last key is there, a partial index update (no key comparison is necessary, all the registries of the index of the last key will be replaced by the index of its sister key) is performed. Then swapping the last key and its sister key is required (unless the last key is the key to be removed from the tree, in which case no further action is necessary, only the size of the tree is decremented and the removal is done).

The purpose of the operations until here is to guarantee that the last key index is not registered anywhere in the tree. Then, the key to be removed from the tree is swapped with the last key. The new last key is dismissed from the tree structure and the size of the tree (*maxIndex*) is decremented by one. All key movements should be accompanied by event ID movements in parallel not to lose the event ID information of the keys.

Just for a comparison, [Fig. 5] depicts a PQ with variable size functionality presented by [1]. To form the tree, 2*N* integer array is used half of which forms the internal nodes while other half forms the leaf nodes. Another auxiliary integer array of size N (this array is named as "Host Leaf" in [Fig. 5]) is used to keep track of which leaf hosts which key index. The procedure to dismiss a key from the tree: Check the two keys registered in the last pair of leaves. Whichever is the winner, change its host leaf information to its current parent node (raise the leaf node). The loser one will replace the key to be dismissed from the tree. Change its host leaf information accordingly and perform a full update. This structure will be used in the comparison section under the name Marin_VS.

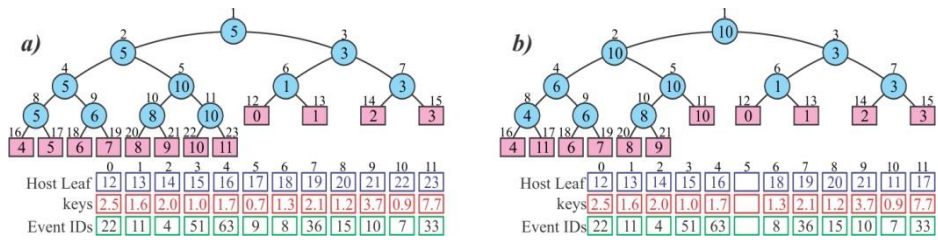

*Figure 5:Variable size CBT proposed by [1] (a) initialized with 12 keys.(b) fifth key is removed from the PQ. 2N auxiliary integer array is used to form the three nodes together with the leaf nodes at the bottom of the tree. The other arrays shown in the figure are Host Leaf, another auxiliary integer array of size N; Keys, the given array hosting the keys; Event IDs, the given integer array containing the information about which key belongs to which event. As a total, 3N auxiliary integer memory space is required for this CBT structure.*

## 4      Comparisons

M. Marín compared various heap, calendar and tree based PQ algorithms together with CBT based ones using the hold model[9] and in addition to using an event-driven particle simulation as a whole[6]. He scaled all the empirical performance results of the tested PQ algorithms to the performance result of his reference PQ. He found that his reference CBT based PQ was the overall fastest PQ candidate.

Here, different PQ algorithms will be compared from two different perspectives; first, the hold model is used in order to assess the default performances of the PQ algorithms (no size variation takes place), second, the candidates are tested for the variable size performances. After the completion of the constant size PQ comparisons, the final states of the PQs are used as the initial state for the variable size PQ comparisons. In the variable size comparison tests, every winner key is dumped from the queue successively until the size of the tree becomes two, much like a sorting algorithm except that the winner keys are not stored anywhere. For the structures that do not have actual size varying functionality, a dummy key value (infinity) is substituted for the winner key values.

Since modern CPUs have a native support for profiling, (e.g. they have a clock register and an instruction for reading it), we can make use of this functionality to measure the amount of time spent for any part of a running code. In order to measure the total time used by any computational block, the CPU clock register is read before and after the evaluated computational block and the difference is calculated. In the case of multiple calls for the same block, the difference can be accumulated in a variable to get the total amount of CPU cycles consumed by that block.

Three different priority distributions were used to generate random key values[9] for the hold model. [Table 1] shows the details of these distributions. PQs were constructed using the given number of keys and the distribution type and an initial $10^6$ hold operations were executed in order to reach a steady state under the given distribution type. Then a loop of $10^6$ hold operations was performed for timing. Timing was achieved by counting the total number of CPU cycles between the beginning and the end of update operations. The accumulated number of CPU cycles was divided by $10^6$ to get an average cost for one hold operation. Following this constant size PQ performance test, the remaining data in the tested structure was used for sorting procedure. A loop of N-2 shrink operations (find the minimum and discard it) was performed. The total number of CPU cycles spent for this loop was registered as the score of the tested structure for the sorting comparison category. The presented empirical results in this article have been scaled to the scores of the implementations based on the same reference CBT structure that Marin used [9] (designated as Marin) so that the performances of the PQs based on the newly proposed structures can be compared to the performances of all other PQs Marín included in his comparisons. The obtained results are presented in two categories: PQ comparisons and sorting comparisons.

A test run for a given number of keys and a distribution type was repeated 10 times but only the averages were used for graphing. For the PQ comparisons, the maximum encountered error (standard deviation divided by average) was 2.6% and for the sorting comparisons, was 3.9%. The computer used for the presented results was a Dell OptiPlex 790 with an Intel Core i5-2400 CPU @3.10 GHz and 8GB RAM. The results obtained using an HP Elite 8300 with an Intel i5 3570 3.4 GHz CPU and 4GB RAM were quite similar to the presented results. The operating systems of both computers were Windows 7 enterprise 64-bit editions. For coding, Visual C++ 2010 programming environment was used. The compilations were done with SSE2 and maximize-speed options enabled.

*Table 1: The Expressions of the three priority distributions used for the hold model. R is a random real value uniformly distributed between 0.0 and 1.0. Expected values of all the distributions are equal to unity.*

| Distribution | Expression for the random priority value. |
|---|---|
| Exponential | -lnR |
| Uniform | 2R |
| Biased | 0.9+0.2R |

## 5  Results and Discussion

[Fig. 6] presents the obtained results in two categories: PQ comparisons and sorting comparisons. The results show that both of the newly introduced structures perform much better than the reference structure. As a data structure to be used for a constant size PQ implementation, they provide about 2X speed-up in the tested range of tree sizes. In a case where PQs continuously shrink, a case very similar to pure sorting, they still perform very well

against the reference CBT. Constant size PQ comparisons show that "Marin_VS" is faster than "Marin" for a certain range of N. This behavior does not show up in referenced article [9]. When the size of the queue is constant, the only difference between the structures "Marin" and "Marin_VS" is the way to find the leaf node for a given key index $i$. The structure "Marin" which is the original CBT structure without node deletions uses the algebraic formula $i+N$, while "Marin_VS" reads it from an array (named Host Leaf in [Fig. 5]). It is interesting that a change in a single statement like this can cause a significant difference in execution time of a computational block. This difference becomes more subtle as the data becomes comparable to the cache size. The reason for this behavior must be hidden inside the internals of modern compilers and the instruction processing of modern CPUs. This behavior disappears in the sorting comparison. This is mainly because when the size variation is imposed "Marin_VS" spends some time to arrange the end of the leaf nodes, whereas "Marin" uses infinity as the new value for the key to be removed and performs a regular update operation.

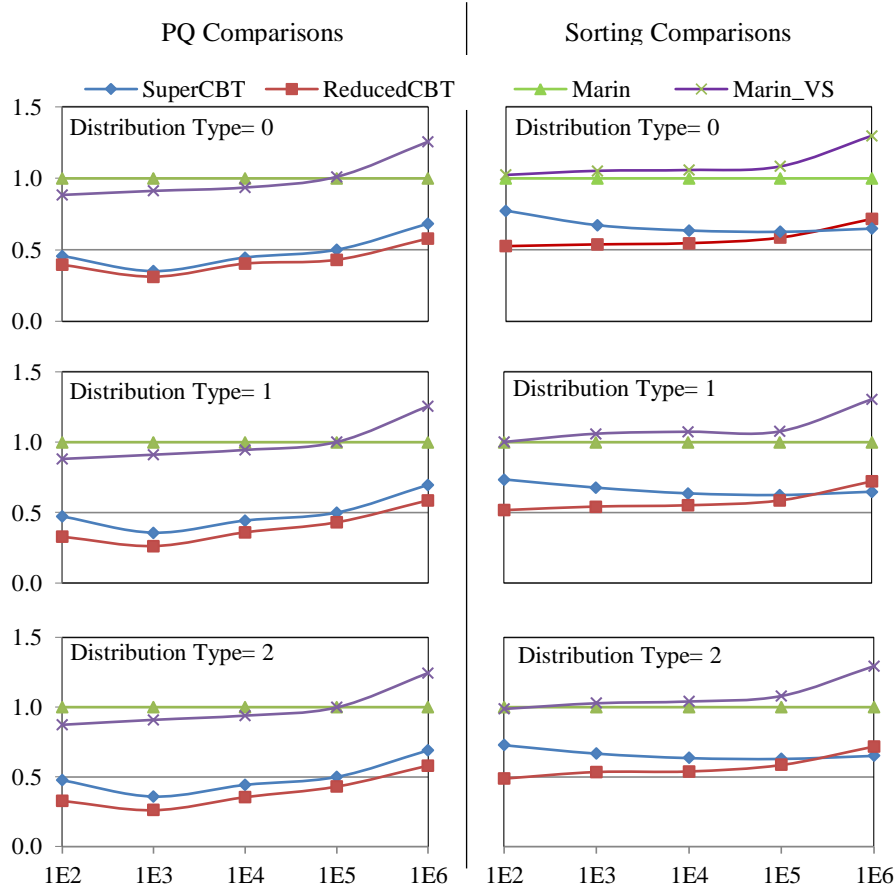

*Figure 6: Empirical comparison results of the tested structures. The legend is at the top of the figure. The left three graphs show the results of constant size PQ implementations while the right three show the results of sorting implementations based on four different tree structures: Marin, the reference CBT structure; Marin_VS, the CBT with variable size functionality; SuperCBT and ReducedCBT, the new structures described in this article. The horizontal axis shows the number of keys, while the vertical axis shows the test scores scaled to the score of the reference structure for the same test. Results for different distribution types have been shown separately although there is not a significant difference between them.*

The sorting comparisons reveal that for large N, SuperCBT scores better than ReducedCBT. This is expected since the number of comparisons for each update operation in a physically shrinking tree goes down as the keys are dismissed from the tree whereas, in the case of ReducedCBT, the keys are not physically removed from the tree, but instead are replaced by infinity and continue to be a part of the tree. Therefore, for a ReducedCBT, the number of comparisons for every update operation is equal to $\log_2 N$ and stays the same.

Another striking fact that can be noticed from the results is that all the CBT structures provide a distribution independent performance. This is because the number of comparisons for update operations is not affected by the key values. The distribution of the key values might only have a secondary effect through the number of node writing operations depending on the results of key comparisons.

ReducedCBT is the fastest and easiest to code among the tested structures in cases when no size variation is required. When size variation functionality is required, depending on the ratio of grow-shrink operations and regular updates, either ReducedCBT or SuperCBT will be the best. In real world applications, the ratio of size varying updates (with an increase or decrease in the number of active keys) and regular updates (without changing the active number of keys) will be case dependent. For example, in a sorting algorithm, all the updates will be size-varying updates. On the other hand, in an EDPS, the majority of collisions are elastic, while some collisions can be inelastic, namely the collision of two particles leading to the formation of a new particle. Such a simulation can be used to study the size distribution of air pollutants [10]. In the case of an elastic collision, the priorities of the involved particles will be updated without changing the number of active particles. However, in the case of an inelastic collision, just one particle will survive into a new body, while the other particle will disappear. Following such an inelastic collision, a regular update plus a size reducing update on the simulation PQ will be necessary. The expected frequency of inelastic collisions will define whether ReducedCBT or SuperCBT is the best structure for the PQ implementation of the simulation.

# 6   Conclusion

ReducedCBT and SuperCBT structures proposed here provide a new way of finding initial parent and sister indices which are needed by the basic tree methods. The novelty in finding the parent and sister indices for the leaves is the incorporation of a computationally intensive method compared to the previously used memory intensive method. The amount of required memory to establish a variable size CBT using the proposed SuperCBT structure is N integers compared to 3N integers required by the previously known variable size CBT structure. Both ReducedCBT and SuperCBT structures lead to faster PQ implementations. A corollary of the proposed SuperCBT structure is that since the end of the key array is emptied after every shrink operation, it is possible to devise an in-place sorting algorithm using this data structure.

By just looking at the properties of the demonstrated CBT structures such as distribution independence, guaranteed $\log_2 N$ comparison per update, usage of an integer array of size N as the only auxiliary memory space, and no recursion; the following general statement can be made. These CBT structures are confirmed to be a natural habitat for researchers who are trying to find the ultimate searching and sorting algorithms.